\documentclass[12pt]{article}
\usepackage{epsfig}
\usepackage{a4,isolatin1}
\usepackage{epsfig,psfrag}

\begin{document}
\begin{center}
{\bf Quantum Effects in Matter-Wave Diffraction\footnote{To appear in the
    Proceedings
    of {\em Quantum Theory and Symmetry}, Cracow, July 2001, edited by
    E. Kapuscik and A. Horzela, World Scientific.}}\\
 \vspace{1cm}
Gerhard C. Hegerfeldt and Thorsten K\"ohler\\
Institut f\"ur Theoretische Physik\\
Universit\"at G\"ottingen\\
\vspace{1cm}
Abstract
\end{center}
\vspace{.2cm}
Advances in micro-technology of the last years have made it possible 
to carry optics textbooks experiments  over to atomic and molecular
beams, such as diffraction by a double
slit or transmission grating. The usual wave-optical approach gives  a
good qualitative 
description of these experiments. However,  small deviations therefrom and
sophisticated  quantum mechanics yield new surprising insights  on the
size of particles and on their interaction with surfaces.

\section {Introduction}
The wave nature of subatomic particles was postulated by de Broglie in
1923 and this idea suffices to explain many diffraction
experiments. Indeed, some people have argued that for matter
diffraction a good optics book is sufficient. The aim of this
contribution is to show that in many cases this is not so, and that
full quantum mechanics may be necessary to describe and evaluate recent more
sophisticated experiments. First of all it is clear from the
statistical interpretation of quantum mechanics that a diffraction
picture is build up slowly from individual particles each of which contributes
a single dot on the screen. A single particle does not give an
interference picture, only the complete particle beam does. This
shows  the wave property belongs to the beam and not the single
particle.

Interestingly, for atoms the simple but fundamental double slit experiment has
been just a thought experiment for a long time. This is due to their
small de Broglie wavelengths $\lambda = H/p$, with $p$ the particle
moment. For usual beam velocities o a few hundred meters per seconds
this is only about 1 {\AA}. Therefore very small slit widths and
distances are needed to obtain observable diffraction angles. Only the
recent advances in micro-technology have made atomic diffraction experiments
possible.

Let us first consider wave optics and a transmission grating of period
$d$ with $N$ slits of widths $s$. If a  classical wave passes through a
grating one observes behind it and outside the original direction an
intensity with characteristic directional modulation. For a perpendicularly
incident plane wave  the diffracted intensity is given by 
\begin{equation}\label{1}
I(\theta) \propto \left(\frac{\sin(\frac{1}{2} N d\, k
  \sin\theta)}{\sin(\frac{1}{2}d\, k\sin \theta)}\right)^2 \times
\left(\frac{\sin(\frac{1}{2}s\, k 
  \sin\theta)}{\frac{1}{2}s\, k  \sin\theta}\right)^2.
\end{equation}
The first factor is the so called grating function, with narrow maxima
of height $Nd$ at the angles 
\begin{equation}
\sin \vartheta_n = n\lambda/d ~~~(n = 0,\pm 1,\pm 2,\cdots)~, 
\end{equation}
and these maxima together with the second factor, the so called slit
function, give the intensity 
\begin{equation}
I_n \equiv I(\theta_n) \propto \frac{\sin^2(n\pi s_0/d)} {(n\pi s_0/d)^2}~.
\end{equation}
For many purposes this simple wave-optical approach gives a
good description of matter diffraction also. However, in some cases
effects from the full quantum theory may be important so that the
simple wave picture is no longer appropriate. This will now be
explained for recent experiments. 

\begin{figure}[tbp]
  \begin{center}
    \epsfig{file=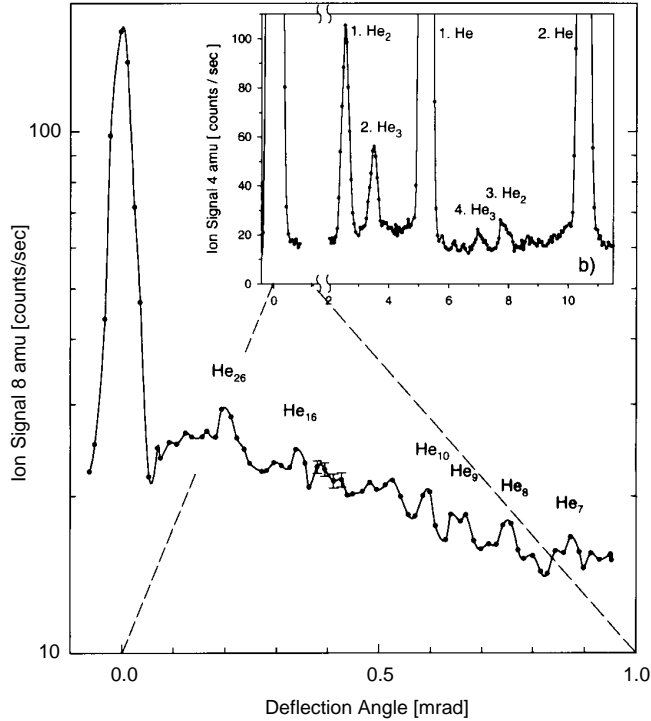
,width=0.6\textwidth}
 \caption{Diffraction pattern of helium atoms and molecules up to
   He$_{26}$ and thereby the first definite detection of these exotic
   molecules (cf. Refs. \cite{Toennies} and
   \cite{Schoellkopf}).}\label{helium26}      
  \end{center}
\end{figure}
The typical experimental setup for atomic and molecular
diffraction  
consists of a beam of particles with very small velocity
distibution which pass through a 
transmission grating or double slit \cite{Pritch,Mlynek,Toennies}. The
 grating used by the Toennies group in G\"ottingen has a period of $d =
 100$ nm.
 A beautiful diffraction 
pattern for a helium beam is shown in Fig. \ref{helium26}. In the
inset in the upper 
right hand corner one observes the first helium diffraction order and
left of it at  half the angle another small maximum. The latter
provided the first direct evidence of the exotic helium molecule
$^4$He$_2$. In the atomic beam there can be helium clusters, all
moving with the same velocity. Therefore their de Broglie wavelengths
and their diffraction angles are inversely proportional to their
mass. The main part of the figure shows diffraction maxima of higher
clusters up to He$_{26}$. The Zeilinger group in Vienna recently
observed diffraction of the fullerenes C$_{60}$ and C$_{70}$ \cite{Zeil}.

Deviations from the simple wave-optical diffraction theory are
expected to occur \cite{HeKoeII} due to 
\begin{itemize}
\item the inner structure of the particles and van der Waals potentials 
\item the spatial extent of the particles
\item the breakup of weakly bound molecules.
\end{itemize}
Are these expected deviations just ``dirt effects'' or do they contain
surprises with useful information? To investigate this question one
needs the full quantum theory.

\section{Quantum Theory of Matter-Wave Diffraction}
\begin{figure}[htbp]
  \begin{center}
\psfrag{Teilchen}{particles}
\psfrag{(Atome)}{(atoms)} 
\psfrag{Gittersteg}{grating bar with}
\psfrag{stark abstossendes Potential}{strongly repulsive potential}
   \epsfig{file=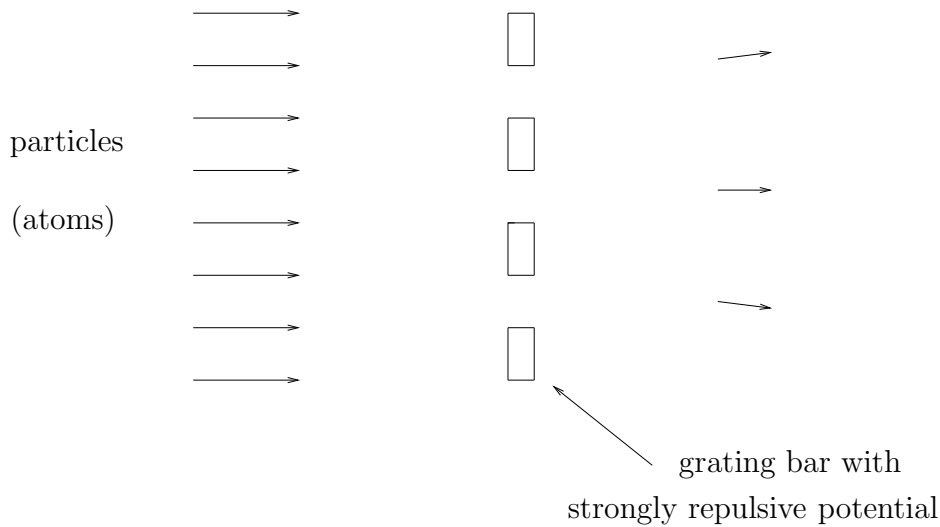,width=0.7\textwidth}
 \caption{Matter diffraction arises from scattering off the grating
   bars.}  \label{qmfig}   
  \end{center}
\end{figure}

First of all one has to realize that matter diffraction off a grating is
not a classical wave phenomenon but a quantum mechanical {\em
  scattering} problem. The diffraction is not caused by the slits but
by scattering of the particles off the grating bars. This is depicted
in Fig. \ref{qmfig}. In addition to the reflective particle-surface
interaction one also has to take a attractive van der Waals surface
interaction into account. 

Starting from the Schr\"odinger equation and using Faddeev scattering
theory in the formulation of Alt, Grassberger and Sandhas
\cite{AGS,HeKoefb} we have 
obtained a general expression for the diffraction intensity $I_n$ of
an extended molecule in the form \cite{prl1}
\begin{equation}
  I_n\propto e^{-(2\pi n\sigma/d)^2}
  \left[
    \frac{\sin^2(n\pi s_{\rm eff}/d)+\sinh^2(n\pi\delta/d)}
    {(n\pi s_{\rm eff}/d)^2 + (n\pi \delta /d)^2 }
  \right]
  \label{diffrint}
\end{equation}
where $s_{\rm eff}$ denotes an effective slit width which is smaller
than $s$. The term $\delta$ diminishes the contrast and the
exponential term takes into account that the number of molecules in
the individual diffraction orders may decrease due to breakups at the
bars and that small variations in the bar widths may occur. All these
parameters may depend on the particle-grating interaction, on  the spatial
extent of the particles and on their velocity.

\subsection{Surface Effects}

\begin{figure}[htbp]
  \begin{center}
   \epsfig{file=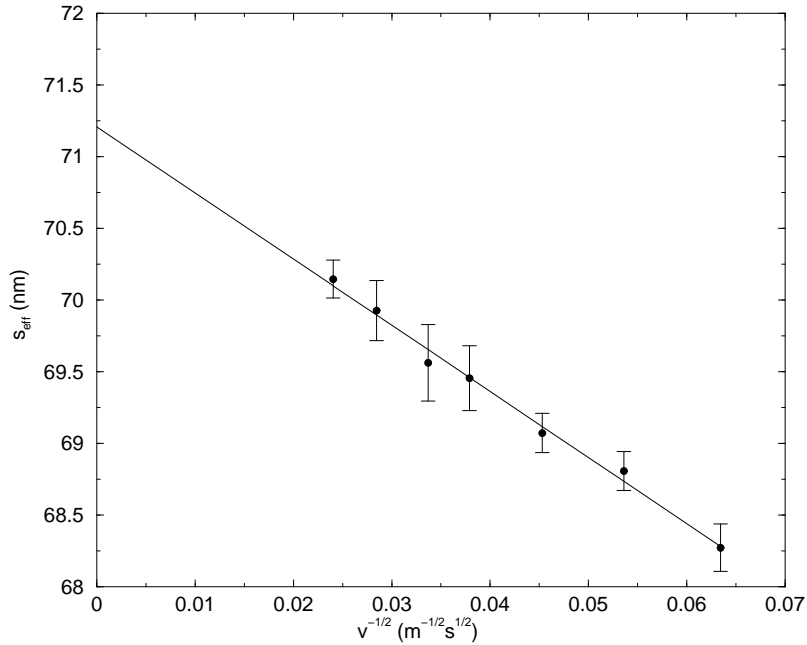,width=0.6\textwidth,angle=270}
 \caption{Plot of $s_{\rm eff}$ over $ 1/\sqrt{v}$ for helium. From
   the slope of   
   the straight line one can calculate $C_3$. The ordinate
   intersection  gives the true slit width of $s = 71.2$ 
   nm.}\label{gerade}        
  \end{center}
\end{figure}

 During the passage  through a slit  an atom experiences an
 additional attractive surface van der Waals potential $V = -C_3/l^3$
 where $l$ denotes the  distance from the surface of a grating bar and
 where $C_3$ depends on the particle species. Quite recently we have
 found the as yet unpublished result that for rare gas atoms $s_{\rm
   eff}$ behaves as $s_{\rm eff}\propto 1/\sqrt{v} $ where $v$ is the
 particle velocity. Therefore we have plotted $s_{\rm
   eff} (v)$ obtained from experimental diffraction patterns for
 variable helium beam velocities as a function of $1/\sqrt{v}$ in
 Fig. \ref{gerade}. 
From the slope one can determine $C_3$. The
 intersection with the ordinate axis yields the true slit width. The
 result agrees with an alternative procedure we have used before
 \cite{prl1}. The method is so sensitive that the geometrical trapeze
 form of the grating base has to be taken into account.

For the evaluation of these measurements one does indeed need quantum
mechanics but qualitatively the difference to wave-optics is not so
pronounced since an optical grating with a dielectric coating can
produce similar effects. This however is no longer true in the
following example.

\subsection{Size of the Helium Dimer}

The main difference to wave-optics occurs for excitations of higher
levels \cite{hekoeprl} as for the helium trimer He$_3$  and for size
and breakup 
effects \cite{HeKoeII}. The latter two effects are particularly
interesting for the exotic 
He$_2$ which is fifty times larger than a hydrogen molecule and whose
binding energy is a hundred million times smaller than that of an electron
in a hydrogen atom. Therefore He$_2$ is extremely fragile and thus 
very difficult to investigate by conventional methods. But at small
diffraction angles most helium dimers arrive at the detectors unbroken
so that deviations of the diffraction pattern from the predictions of
wave-optics can be measured and analyzed.

In Ref. \cite{prl2} experimental dimer diffraction intensities up to
7th order have been fitted to the   expression $I_n$ from
Eq. (\ref{diffrint}) and  $s_{\rm eff}$ was determined. By quantum mechanical
multi-channel scattering theory we obtained a relation between $s_{\rm
  eff}(v)$ and the inter-nuclear distance $\langle
r \rangle$. A simplified result is
$\langle
r \rangle/2 = s_{\rm eff}^{\rm He}(v)-s_{\rm eff}^{^4{\rm He}_2}(v)$, which gives $<r> \approx 50$ \AA{}. The more precise
theory yields $\langle
r \rangle = 52 \pm 4$ \AA{}.

\section{Conclusions}

These examples show that refinements of a simple textbook experiment on matter
diffraction can lead to new quantum mechanical applications in atomic
and molecular physics. In particular we have discussed 
\begin{itemize}
\item its use as a quantum mass spectrometer 
\item exploitation of quantum effects for 
\begin{itemize}
\item particle surface interaction
\item influence of the particle size on the diffraction pattern:
size determination
\item detection of excited energy levels for He$_3$ (work in progress)
\end{itemize}
\end{itemize}

As a further development we mention the setup of an atom
interferometer which, in a first approximation, can be understood
similarly as in wave-optics, but to describe finer effects  one  needs
full quantum mechanics.


\begin{thebibliography}{99}
\bibitem{Pritch}D.~W.~Keith, M.~L.~Schattenburg, H.~I.~Smith, D.~E.~Pritchard, 
  Phys.~Rev.~Lett.~{\bf 61}, 1580 (1988)
\bibitem{Mlynek}O.~Carnal and J.~Mlynek, Phys.~Rev.~Lett.~{\bf 66}, 2689 (1991)
\bibitem{Toennies}W.~Sch\"ollkopf, J.~P.~Toennies, Science {\bf 266},
  1345 (1994) 
\bibitem{Schoellkopf}W.~Sch\"ollkopf, Doctoral Dissertation, University
  of G\"ottingen (1998)
\bibitem{Zeil} M. Arndt, O. Nairz, J. Vos-Andreae, C. Keller, G. van
  der Zouv, and A. Zeilinger, Nature (London), {\bf 401}, 680 (1999)
\bibitem{HeKoeII}
  G.~C.~Hegerfeldt, T.~K\"ohler, Phys.~Rev.~A {\bf 57}, 2021 (1998); 
  Phys.~Rev.~A {\bf 61}, 023606 (2000)
\bibitem{AGS} E.~O.~Alt, P.~Grassberger, and W.~Sandhas, Nucl.~Phys.~B {\bf 2},
  167 (1967)
\bibitem{HeKoefb}
  G.~C.~Hegerfeldt, T.~K\"ohler, Few Body Systems Suppl. {\bf 10}, 263 (1999)
\bibitem{prl1}R.~E.~Grisenti, W.~Sch\"ollkopf, J.~P.~Toennies,
  G.~C.~Hegerfeldt,   T.~K\"ohler, Phys.~Rev.~Lett.~{\bf 83}, 1755 (1999)
\bibitem{hekoeprl}G.~C.~Hegerfeldt, T.~K\"ohler, Phys.~Rev.~Lett. {\bf 84},
3215 (2000) 
\bibitem{prl2}R.~E.~Grisenti, W.~Sch\"ollkopf, J.~P.~Toennies,
  G.~C.~Hegerfeldt,   T.~K\"ohler, M.~Stoll, Phys.~Rev.~Lett. {\bf
    85}, 2284 (2000) 

\end{thebibliography}
\end{document}